\documentclass[aps,floats,twocolumn,pre,10pt,superscriptaddress]{revtex4-1}
\usepackage{amsmath,amssymb,amsthm,amsfonts,bm,bbm,braket,enumitem,mathtools}
\usepackage[english]{babel}

\usepackage{natbib}
\usepackage{hyperref}
\hypersetup{
	pdfauthor={Cosimo Lupo, Giorgio Parisi and Federico Ricci-Tersenghi},
	pdftitle={The random field XY model shows replica symmetry breaking and marginally unstable ferromagnetism},
	colorlinks,
	linktocpage=true,
	pdfstartpage=1,
	pdfstartview=FitV,
	breaklinks=true,
    pageanchor=true,
    pdfpagemode=UseOutlines,
    plainpages=false,
    bookmarksnumbered,
    bookmarksopen=true,
    bookmarksopenlevel=1,
    hypertexnames=true,
    pdfhighlight=/O,
    urlcolor=blue,
    linkcolor=blue,
    citecolor=blue,
}

\allowdisplaybreaks

\usepackage[utf8x]{inputenc}	
\usepackage[T1]{fontenc} 	

\usepackage{graphicx}

\begin{document}

\title{The random field XY model on sparse random graphs shows\\
replica symmetry breaking and marginally stable ferromagnetism}

\author{Cosimo Lupo}
\email[Corresponding author: ]{cosimo.lupo89@gmail.com}
\affiliation{Laboratoire de Physique de l'\'Ecole Normale Sup\'erieure (LPENS), ENS, Universit\'e PSL, CNRS, Sorbonne Universit\'e, Universit\'e Paris-Diderot, Sorbonne Paris Cit\'e, 75005 Paris, France}
\author{Giorgio Parisi}
\affiliation{Dipartimento di Fisica, Sapienza Universit\`a di Roma \& INFN, Sezione di Roma1 \& CNR-Nanotec, Rome unit, 00185 Rome, Italy}
\author{Federico Ricci-Tersenghi}
\affiliation{Dipartimento di Fisica, Sapienza Universit\`a di Roma \& INFN, Sezione di Roma1 \& CNR-Nanotec, Rome unit, 00185 Rome, Italy}
\date{\today}

\begin{abstract}
The ferromagnetic XY model on sparse random graphs in a randomly oriented field is analyzed via the belief propagation algorithm. At variance with the fully connected case and with the random field Ising model on the same topology, we find strong evidences of a tiny region with Replica Symmetry Breaking (RSB) in the limit of very low temperatures. This RSB phase is robust against different choices of the external field direction, while it rapidly vanishes when increasing the graph mean degree, the temperature or the directional bias in the external field.  The crucial ingredients to have such a RSB phase seem to be the continuous nature of vector spins, mostly preserved by the $\mathrm{O}(2)$-invariant random field, and the strong spatial heterogeneity, due to graph sparsity. We also uncover that the ferromagnetic phase can be marginally stable despite the presence of the random field. Finally, we study the proper correlation functions approaching the critical points to identify the ones that become more critical.
\end{abstract}

\maketitle

\section{Introduction}

Ferromagnetic systems in random magnetic fields are the archetypal of disordered systems, representing the simplest way to introduce a certain degree of \textit{quenched} disorder into an ordered substrate. The most famous among them, the Random Field Ising Model (RFIM), has been already introduced more than forty years ago by Larkin~\cite{Larkin1970} in the context of superconductors. From then, the RFIM has been the object of countless speculative studies~\cite{ImryMa1975, SchneiderPytte1977, BricmontKupiainen1987, TissierTarjus2011, FytasMartinMayor2013, LucibelloEtAl2014}, as well as applied in a huge number of fields, from condensed matter~\cite{FishmanAharony1979} to strongly correlated electronic systems~\cite{EfrosShklovskii1975, KirkpatrickBelitz1994, Dagotto2005}, from liquids and colloid mixtures~\cite{deGennes1984,VinkEtAl2006} to opinion dynamics and social interactions~\cite{Galam1997, MichardBouchaud2005}.

Despite their effectiveness in numerous applications, random field systems still lack of a general theoretical framework that takes into account their static and dynamic properties. In particular, two issues on the RFIM have generated a long debate over the years: the correctness of the dimensional reduction by Parisi and Sourlas~\cite{ParisiSourlas1979} and the presence of a glassy phase. In particular, the latter debate has finally found a definitive answer only a few years ago, thanks to the argument according to which a Replica Symmetry Broken (RSB) phase can not be observed for the RFIM on any topology~\cite{KrzakalaEtAl2010, KrzakalaEtAl2011}, provided the system is at thermal equilibrium. Indeed, given the positive couplings of the model, connected correlation functions are always nonnegative and hence spin glass susceptibility is always upper bounded by the ferromagnetic one. This crucial observation rules out the possibility of a proper spin glass phase, while nothing can be deducted from it about the behaviour of the RFIM exactly at the critical point, where at variance some evidences of the presence of many states in the Gibbs measure~\cite{PeruginiRicciTersenghi2018} have been recently found.

A natural extension of the previous argument from scalar to vector spins has not yet been accomplished, and the reason is that it is not possible. Indeed, when taking into account small fluctuations of vector spins around their equilibrium configurations, negative effective correlations could in principle take place thanks to the small-fluctuation mechanism of vector spins, even if all the couplings are ferromagnetic. So the argument of Ref.~\cite{KrzakalaEtAl2010} can not be applied for vector spins.

At this point, we wonder if these low-energy excitations are enough to break the symmetry between replicas and hence yield a RSB phase for some suitable vector spin models with ferromagnetic couplings in a random field. This is the main aim of this work. We will focus on the simplest magnetic model with continuous variables, the XY model, where spins are $m=2$\,-\,component vectors of unit norm. Since in the fully connected case there is no chance of getting RSB when the system is purely ferromagnetic, we move to the case of sparse random graphs, where the heterogeneity provided by the sparsity can enhance the disordering effect of the random field.

If on one hand the local treelike topology of these graphs will allow us to exactly solve the model via Belief-Propagation (BP) algorithms, on the other hand we will also be able to analyze the features of the typical low-energy excitations of the system, exploiting the tools already developed when characterizing the instabilities of the spin glass XY model in an external field~\cite{LupoRicciTersenghi2018}. Moreover, we will mostly focus on the zero-temperature behaviour of the XY model, recalling and extending the discretization arguments already presented in~\cite{LupoRicciTersenghi2017}. Finally, we will complete the description of the XY model on diluted graphs in a random field by defining and studying suitable correlation functions, identifying their behavior in each phase. The critical points of the model will be naturally interpreted via the divergence of a suitable correlation length, so recognizing the kinds of correlations that become critical in each case.

Notice that traces of a glassy behaviour for a ferromagnetic XY model in a random field have been already found on $d=2$ regular lattices, despite additional hyphoteses were added to the original model in order to prevent other kinds of long-range order --- as the Berezinski{\u{\i}}-Kosterlitz-Thouless phase~\cite{Berezinskii1971, KosterlitzThouless1973} ---, e.\,g. by removing the $2\pi$-periodicity of the spins~\cite{LeDoussalGiamarchi1995} or by changing the periodicity of the external random field~\cite{CardyOstlund1982}. In our model, instead, it is the random topology that prevents the appearance of vortexes and spin waves, while spins are the usual $2\pi$-periodic ones in a random external field with the same periodicity.

This paper is organized as follows. In section~\ref{sec:model} we introduce the XY model and we briefly recap the belief-propagation algorithm, that will allow us to solve it on sparse random graphs. In section~\ref{sec:RS_sol} we will then write down and solve the corresponding BP equations, drawing the entire field vs temperature phase diagram and actually finding a RS unstable region in the very low-temperature regime. In section~\ref{sec:RSB_phase} we will comment the nature of this RS unstable region, checking its robustness and trying to better characterize the role of small fluctuations that invalidate the argument preventing RSB in the RFIM. The expected connection between a second-order phase transition and the divergence of some correlation length at the critical point is finally attained in section~\ref{sec:corr_funcs}. Final remarks are exposed in sec.~\ref{sec:conclusions}.

\section{The XY model}
\label{sec:model}

The simplest magnetic model with continuous variables is represented by the XY model, where spins $\boldsymbol{\sigma}_i$'s are unit vectors lying in the $xy$ plane. Due to the normalization constraint, there is just a unique degree of freedom per spin, that can be effectively described by an angle $\theta_i\in[0,2\pi)$. The external field is a $m=2$\,-\,dimensional vector as well, with modulus $H_i$ and direction $\phi_i\in[0,2\pi)$. Given a purely ferromagnetic interaction of strength $J>0$ between each couple of spins --- corresponding to undirected edges $(i,j)$'s in the edge set~$\mathcal{E}$ of the interacting graph~$G$ ---, the random field XY model is ruled by the following Hamiltonian:
\begin{equation}
	\mathcal{H}[\{\theta_i\}]=-J\sum_{(i,j)\in\mathcal{E}}\cos{(\theta_i-\theta_j)} -\sum_i H_i\cos{(\theta_i-\phi_i)}
	\label{eq:H_XY}
\end{equation}

In a previous work of ours~\cite{LupoRicciTersenghi2017} we showed that in the low-temperature region vector spin glass models are by far more unstable toward glassiness than scalar models, due to the possibility of small fluctuations around equilibrium configurations of spins. This continues to be true in presence of an external field, provided its randomness in the local direction $\phi_i$ of the field~\cite{LupoRicciTersenghi2018}, as also pointed out for the first time by Sharma and Young~\cite{SharmaYoung2010}. So let us assume $\phi_i$ being a random variable drawn from a suitable probability distribution~$\mathbb{P}_{\phi}$. In order to enhance the most the disordering effect of the field, we take $\mathbb{P}_{\phi}(\phi_i)=1/2\pi$ over the $[0,2\pi)$ interval, while $H_i$ can be set equal to $H$ on each site without any loss of generality~\cite{LupoRicciTersenghi2018}.

In order to exactly solve the random field XY model, we assume the underlying graph~$G$ to be drawn from the ensemble of Random Regular Graphs (RRGs) of fixed connectivity $C$, also generically referred to as Bethe lattices~\cite{Book_JansonEtAl2000, Book_Bollobas2001}. In this way, due to the local treelike structure of such graphs, Bethe approximation~\cite{Bethe1935} successfully leads to the replica symmetric solution via the belief propagation algorithm~\cite{Book_MezardMontanari2009}, equivalent to the replica symmetric cavity method~\cite{MezardParisi2001, MezardParisi2003}.

\section{RS solution via belief propagation}
\label{sec:RS_sol}

\subsection*{BP Equations and the Population Dynamics Algorithm}

The basic object to look at in the belief-propagation algorithm is --- for each directed edge $i\to j$ --- the probability distribution $\eta_{i\to j}(\theta_i)$ of the angle $\theta_i$ in the modified graph without such edge, also known as \textit{cavity message}. Indeed, these objects allow to compute one-point and two-point probability distributions
\begin{subequations}
	\begin{equation}
	\begin{split}
		\eta_i(\theta_i) &= \frac{1}{\mathcal{Z}_i}\,e^{\,\beta H\cos{(\theta_i-\phi_i)}}\\
		&\qquad\times\prod_{k\in\partial i}\int d\theta_k\,e^{\,\beta J\cos{(\theta_i-\theta_k)}}\,\eta_{k\to i}(\theta_k)
		\label{eq:BP_XY_onePoint}
	\end{split}
	\end{equation}
	\begin{equation}
		\eta_{ij}(\theta_i,\theta_j) = \frac{1}{\mathcal{Z}_{ij}}\,e^{\,\beta J\cos{(\theta_i-\theta_j)}}\,\eta_{i\to j}(\theta_i)\,\eta_{j\to i}(\theta_j)
		\label{eq:BP_XY_twoPoints}
	\end{equation}
\end{subequations}
and from them any physical observable in the Bethe approximation~\cite{Book_MezardMontanari2009}.

Cavity messages $\eta_{i\to j}$'s satisfy a set of recursive relations, widely known as cavity equations or better belief-propagation equations~\cite{YedidiaEtAl2003}:
\begin{equation}
\begin{split}
	\eta_{i\to j}(\theta_i) &\equiv \mathcal{F}[\{\eta_{k\to i}\},J,\phi_i]\\
	&= \frac{1}{\mathcal{Z}_{i\to j}}\,e^{\,\beta H\cos{(\theta_i-\phi_i)}}\\
	&\qquad\times\prod_{k\in\partial i\setminus j}\int d\theta_k\,e^{\,\beta J\cos{(\theta_i-\theta_k)}}\,\eta_{k\to i}(\theta_k)
	\label{eq:BP_XY_eqs}
\end{split}
\end{equation}

These equations can be numerically solved on a given instance of the problem --- namely on a fixed graph $G$ with a given set of field directions $\phi_i$'s --- or directly in a \textit{disorder-averaged} frame, provided by the Population Dynamics Algorithm (PDA)~\cite{AbouChacraEtAl1973}. This approach focuses on the probability distribution of the cavity messages $P[\eta_{i\to j}]$ as if they were random variables, actually solving the distributional version of BP equations~(\ref{eq:BP_XY_eqs}):
\begin{equation}
\begin{split}
	P[\eta_{i\to j}] &= \mathbb{E}_{G,\phi} \int\prod_{k=1}^{C-1}\mathcal{D}\eta_{k\to i}\,P[\eta_{k\to i}]\\
	&\qquad\times\delta\Bigl[\eta_{i\to j}-\mathcal{F}[\{\eta_{k\to i}\},J,\phi_i]\Bigr]
	\label{eq:BP_XY_eqs_distr}
\end{split}
\end{equation}
In this way, the fixed point $\{\eta^*_{i\to j}\}$ of BP equations~(\ref{eq:BP_XY_eqs}) on a given graph $G$ is substituted by their fixed-point probability distribution $P^*[\eta_{i\to j}]$ over the whole graph ensemble.

Equations~(\ref{eq:BP_XY_eqs}) can be numerically solved by reproducing $P[\eta_{i\to j}]$ via a set of $\mathcal{N}$ cavity messages, with a very mild dependence on the size~$\mathcal{N}$ of the population, such that the thermodynamic limit can be easily attained. Indeed, distributional version of BP equations are said to solve the model on an \textit{infinite} tree.

A crucial issue in these numerical simulations is how to deal with probability distributions over continuous variables. If these functions are particularly regular, e.\,g. $2\pi$\,-\,periodic, then it is convenient to project them onto orthonormal polynomials, as actually done for the spin glass XY model at zero field~\cite{CoolenEtAl2005, SkantzosEtAl2005, MarruzzoLeuzzi2015, LupoRicciTersenghi2017}; but when the function is not so smooth it may be convenient to discretize it~\cite{MezardEtAl2011}. Due to the presence of the external field that does not allow a Fourier expansion around the uniform solution $\eta_{i\to j}(\theta_i)=1/2\pi$ as in Ref.~\cite{LupoRicciTersenghi2017}, we follow the second approach, dividing the $[0,2\pi)$ interval into $Q$ equal bins and hence moving from the XY model to the \textit{$Q$-state clock model}~\cite{NobreSherrington1986, IlkerBerker2013, IlkerBerker2014, MarruzzoLeuzzi2015, LupoRicciTersenghi2017, CaglarBerker2017a, CaglarBerker2017b}.

Even though seeming inefficient, in fact we showed that the $Q$-state clock model represents both an efficient and reliable proxy for the spin glass XY model, with an error going to zero exponentially fast in~$Q$ both at finite and zero temperature~\cite{LupoRicciTersenghi2017}. Indeed, the convergence is actually boosted by the disordering effect provided by the quenched disorder. In this case, however, because of the polarizing effect induced by the ferromagnetic couplings, in the zero-temperature limit the $Q$-dependence could get worse, hence we will have to carefully deal with it.

Finally, we redirect the reader to Refs.~\cite{LupoRicciTersenghi2017, LupoRicciTersenghi2018, Thesis_Lupo2017} for further details about the numerical solution of the XY model on Bethe lattices via the $Q$-state clock model, in particular for what regards the involved zero-temperature limit, the analysis of the linear stability and the arise of metastable solutions when an external field is present.

\subsection*{RS Solution}

By exploiting the PDA with $\mathcal{N}=10^6$ on a $Q$-state clock model with $Q=64$, we can actually find the replica symmetric solution of the ferromagnetic XY model in a random field at finite temperature, provided that finite-$Q$ corrections can be already neglected in this regime~\cite{LupoRicciTersenghi2017}. Notice also that, despite dealing with the discretized version of the initial model, the distribution $\mathbb{P}_{\phi}$ of the field direction can still be considered as continuously defined over the $[0,2\pi)$ interval.

Large values of the temperature $T$ and of the field intensity $H$ yield the paramagnetic solution, which of course is no longer the uniform one over the $[0,2\pi)$ interval, due to the presence of the random field. Then, when getting closer to the small-field -- small-temperature region, such solution becomes unstable and an ordered phase takes place, characterized by a nonvanishing global magnetization continuously growing from zero when trespassing the critical point. So far, the picture is similar to usual random field models, as e.\,g. the Ising model on Bethe lattices~\cite{MoroneEtAl2014}. In particular, in the zero-field limit such instability line matches the instability point on the~$T$ axis given by the analytic condition
\begin{equation}
	\frac{I_1(\beta_c J)}{I_0(\beta_c J)} = \frac{1}{C-1}
\end{equation}
for a $C$-RRG, as derived in Refs.~\cite{CoolenEtAl2005, SkantzosEtAl2005, MarruzzoLeuzzi2015, LupoRicciTersenghi2017}.

The exact location of the second-order critical line between paramagnetic and ferromagnetic phases can be found via the Susceptibility Propagation approach~\cite{LupoRicciTersenghi2017, LupoRicciTersenghi2018}, namely the stability analysis of the BP fixed point $\mathbb{P}^*[\eta_{i\to j}]$. In more detail, the perturbations $\{\delta\eta_{i\to j}\}$ of the fixed-point cavity marginals $\{\eta^*_{i\to j}\}$ evolve according to the linearized version of the BP equations~(\ref{eq:BP_XY_eqs}), with a global growth rate $\lambda_{\text{BP}}$ then defined as
\begin{equation}
	\lambda_{\text{BP}} \equiv \lim_{t\to\infty}\frac{1}{t\,\mathcal{N}}\sum_{(i\to j)}\ln\int |\delta\eta_{i\to j}(\theta)| d\theta 
\end{equation}
and evaluated over the $\mathcal{N}$-sized population of cavity messages and related perturbations. Stability is hence highlighted by a negative $\lambda_{\text{BP}}$, with the critical line then identified by the condition $\lambda_{\text{BP}}=0$. Notice also that taking the average of the logarithm of norm perturbations means considering their typical value, which indeed is what rules the stability of the BP fixed point.

\subsection*{The zero-temperature limit}

When working directly at $T=0$, it is the case to move to the large-deviation representation of the cavity messages, $\eta_{i\to j}(\theta_i) \equiv \exp{\{\beta h_{i\to j}(\theta_i)\}}$, where cavity fields $h_{i\to j}$'s are normalized up to an additive constant so to have maximum at zero height and satisfy the following zero-temperature BP equations~\cite{LupoRicciTersenghi2017, Thesis_Lupo2017}:
\begin{equation}
\begin{split}
	h_{i\to j}(\theta_i) &\equiv \mathcal{F}_0[\{h_{k\to i}\}, J,\phi_i]\\
	&\cong H\cos{(\theta_i-\phi_i)}\\
	&\qquad+\sum_{k\in\partial i\setminus j}\max_{\theta_k}{\left[h_{k\to i}(\theta_k)+J\cos{(\theta_i-\theta_k)}\right]}
	\label{eq:BP_XY_eqs_zeroTemp}
\end{split}
\end{equation}

Once again, linear stability can be studied by looking at the global growth rate $\lambda_{\text{BP}}$ of perturbations, $\delta h_{i\to j}$'s, that evolve according to the linearized version of Eqs.~(\ref{eq:BP_XY_eqs_zeroTemp}) and are additively normalized so to cross the zero in correspondence of the argmax $\theta^*_i$ of the corresponding cavity field $h_{i\to j}$ (indeed, it would represent the first-order term of the $1/\beta$ expansion).

Numerically, the discrete nature of the $Q$-state clock model poses a problem in the evolution of the perturbations, that in principle would behave as those of a truly discrete model (e.\,g. the Ising model). To restore the continuous nature of perturbations and hence of the model numerically simulated, it is enough to evaluate argmax's in the linearized zero-temperature BP equations over the real values rather than picking from the $Q$-discretized subset, through a second-order polynomial interpolation of the cavity fields and of their perturbations. Otherwise, without this precaution, all the perturbations will eventually collapse identically to zero, instead of being continuously enlarged or shrank at each iteration of BP. Hence, the wrong handling of zero-temperature perturbations leads to a completely wrong analysis of the stability, e.\,g. resulting in a macroscopically wrong estimate of the critical point.

\subsection*{Extrapolation in Q}

Let us now actually solve the BP equations. We start directly from the $T=0$ case, looking for the BP fixed point separately for each value of the field modulus, eventually averaging the resulting physical observables over $\mathfrak{r}=5$ independent runs. In this way, we are sure to deal with a numerical precision high enough to clearly detect any possible phase transition.

\begin{figure}[!t]
	\centering
	\includegraphics[width=\columnwidth]{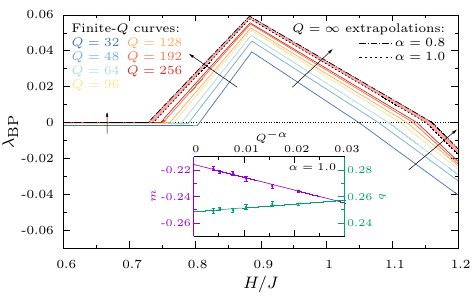}
	\caption{\textit{Main plot}: Stability parameter $\lambda_{\text{BP}}$ for the ferromagnetic XY model in a randomly oriented field of fixed intensity $H$ on a $C=3$ RRG, at $T=0$. The $Q$-dependence is not negligible, leading to an increase of width of the RSB region in the $Q\to\infty$ limit. In the region of interest, the $\lambda_{\text{BP}}(H)$ curve can be reliably described by a piecewise linear function. The extrapolation is then performed on the two parameters (slope $m$ and intercept $q$) for each of the four intervals, following a power-law, $Q^{-\alpha}$, with $\alpha$ reliably belonging to the $[0.8,1.0]$ interval. The corresponding region of extrapolation is highlighted in light red between the two black curves. \textit{Inset}: The power-law $Q\to\infty$ extrapolation performed on the third linear interval (the one from $H/J \sim 0.9$ to $H/J \sim 1.10$), plotted for $\alpha=1.0$ as reference. The left y axis refers to the scale for the slope $m$, the right y axis to the scale for the intercept $q$. The represented data-points go from $Q=48$ (the rightmost one) to $Q=256$ (the leftmost one), while the data-point for $Q=32$ has been excluded from the fit for both $m$ and $q$.}
	\label{fig:lambdaBP_XYRF_C03_T0_Qextrap}
\end{figure}

The careful analysis of the stability of BP fixed points so obtained allowed us to uncover a tiny region of instability with respect to both the paramagnetic and the ferromagnetic solutions, located exactly in-between them. In Fig.~\ref{fig:lambdaBP_XYRF_C03_T0_Qextrap} we reported the $C=3$ case, where such instability region can be well appreciated. Let us refer to the two critical values of the field modulus for the $P-SG$ and $SG-F$ phase transitions as $H^{(+)}_c$ and $H^{(-)}_c$, respectively. Despite the necessary interpolation over the real-valued angles in the zero-temperature algorithm --- that allowed us to recover the continuous nature of perturbations --- it is clear from the figure that a residual dependence on~$Q$ is still present for the $\lambda_{\text{BP}}(H)$ curves, and hence an explicit $Q\to\infty$ extrapolation is needed in order to recover the XY values of $H^{(+)}_c$ and $H^{(-)}_c$. Indeed, it is evident that the larger $Q$, the wider $\Delta H_c$, namely the width of the RSB region.

In order to extrapolate the correct value for the XY model, we divided the region of interest in four intervals, where the $\lambda_{\text{BP}}(H)$ curve can be reliably approximated as a linear function. Then, for each interval, the two parameters $m$ (the slope) and $q$ (the intercept) of the linear approximation seem to suffer from power-law corrections with respect to their $Q\to\infty$ limit:
\begin{equation}
	m^{(Q)} \sim m^{(\infty)} + \frac{a}{Q^{\alpha}} \quad , \quad q^{(Q)} \sim q^{(\infty)} + \frac{b}{Q^{\alpha}}
	\label{eq:Q_extrap}
\end{equation}
with $\alpha$ that can be reliably assigned to the $[0.8,1.0]$ interval (inset of Fig.~\ref{fig:lambdaBP_XYRF_C03_T0_Qextrap}).

Since the range of $Q$ values does not span many orders of magnitude, the $Q$-dependence of $m$ and $q$ could be even fitted with an exponential function, with a characteristic scale $Q^* \sim \mathcal{O}(10^2)$. However, the resulting $Q\to\infty$ extrapolation is consistent with the power-law one shown above, within the confidence interval about the $\alpha$ exponent.

These power-law corrections could seem as in contrast with the conclusions of Ref.~\cite{LupoRicciTersenghi2017}, where an exponential convergence of the physical observables was observed, with a characteristic scale $Q^* \sim \mathcal{O}(1)$. In fact, there are two main points to be taken into account: \textit{i)} as already stated in the aforementioned work, the convergence of physical observables is still exponential in~$Q$ even at $T=0$, provided there is an ``enough'' disordering action by the quenched couplings, and here it is not the case; otherwise, when the ordering effect of the (ferromagnetic) couplings prevails, it has been already argued that the convergence could have dramatically slowed down to a power-law (i.\,e., the characteristic scale of the exponential decay could have diverged) ; \textit{ii)} in the same work, we also observed that some critical lines used to converge dramatically fast in~$Q$, as the one between the paramagnetic and the spin glass phases, while other did not, as the one between the RS ferromagnetic and the mixed RSB phase close to the zero-temperature and ``ordered model'' axes; again, it is due to the combined ordering effect of both the temperature and the couplings.

The $Q\to\infty$ extrapolated values for the two critical fields are reported in Tab.~\ref{tab:Tc_vs_C} for many values of the connectivity $C$ of the RRG graphs, together with the amplitude $\Delta H_c$ of the RSB region at $T=0$. We rescaled the couplings of $1/(C-1)$ so to get finite $C\to\infty$ results. As a consequence of this, also $H$ and $T$ get rescaled by the same amount, namely $1/(C-1)$, with respect to the usual $J=1$ choice. Within the numerical precision we attained --- mainly due to the systematic error on the choice of $\alpha$, rather than to the statistical error on the single measures --- we are confident that the width $\Delta H_c$ of the RSB region shrinks to zero linearly in $C$, likely going to zero in the $(7,8)$ interval. However, an exponentially-small (or power-law) RSB region cannot be in principle ruled out for $C \gtrsim 8$, given the difficulty to further enhance the numerical precision we obtained. What is sure, is that exactly in the fully-connected (or SK) limit, there is no RSB region at all --- as well known in the literature and also shown here in App.~\ref{app:SK_limit} --- and $H^{(+)}_c=H^{(-)}_c=1/2$.

\begin{table}[t]
	\caption{End-points of the RSB phase in the phase diagram of the ferromagnetic XY model in a random field on random $C$-regular graphs. The coupling strength $J=1/(C-1)$ has been chosen such to have a finite value for $\lim_{C\to\infty} T_c(H=0) = 1/2$, as well as for $\lim_{C\to\infty} H_c(T=0) = 1/2$. From $C=5$ the RSB area is no longer detectable with finite-$T$ numerics, while it disappears for $C \gtrsim 8$ even from the $T=0$ results, with $H^{(-)}_c$ and $H^{(+)}_c$ coinciding within the numerical precision used. The last line refers to the SK analytic results from App.~\ref{app:SK_limit}.}
	\begin{tabular}{c|c|c|c|c}
			$C$ & $H^{(-)}_c$ & $H^{(+)}_c$ & $\Delta\,H_c $ & $T^*$\\
		\hline
			3 & 0.367(3) & 0.578(3) & 0.211(4) & 0.012(1)\\
			4 & 0.503(3) & 0.667(4) & 0.164(5) & 0.004(1)\\
			5 & 0.560(3) & 0.667(4) & 0.107(5) & /\\
			6 & 0.578(4) & 0.646(4) & 0.068(6) & /\\
			7 & 0.577(4) & 0.611(5) & 0.033(6) & /\\
			8 & 0.575(5) & 0.575(5) & 0.000(6) & /\\
			12 & 0.554(5) & 0.554(5) & 0.000(6) & /\\
			16 & 0.539(5) & 0.539(5) & 0.000(6) & /\\
			20 & 0.530(5) & 0.530(5) & 0.000(6) & /\\
			\dots & \dots & \dots & \dots & \dots\\
			$\infty$ & 1/2 & 1/2 & 0 & 0\\
	\end{tabular}
	\label{tab:Tc_vs_C}
\end{table}

For $H<H_c^{(-)}$ we observe that $\lambda_{\text{BP}}$ is constant and slightly negative for any finite value of $Q$. In the $Q\to\infty$ limit it becomes zero, signaling a marginal ferromagnetic phase. We will discuss below in more detail this ferromagnetic marginal phase.

\begin{figure}[!t]
	\centering
	\includegraphics[width=\columnwidth]{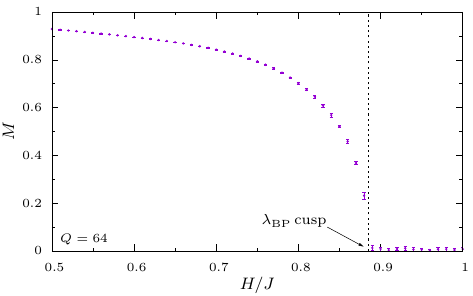}
	\caption{Modulus $M$ of the global magnetization as a function of the field modulus $H$ for the ferromagnetic XY model in a randomly oriented field of fixed intensity $H$ at $T=0$. The underlying topology is again the $C=3$ RRG ensemble. $M$~drops continuously to zero in correspondence of the cusp in the $\lambda_{\text{BP}}(H)$ curve, signaling the phase transition between the magnetized mixed phase (on the left) and the unmagnetized spin glass phase (on the right), both RS-unstable. The datapoints refer to $Q=64$, and the cusp location is coherent with this choice. The $Q\to\infty$ extrapolation does not qualitatively change the picture, only resulting in a slight shift of the cusp (see Fig.~\ref{fig:lambdaBP_XYRF_C03_T0_Qextrap}).}
	\label{fig:magCurve_XYRF_T0_C03}
\end{figure}

One may also wonder why $\lambda_\text{BP}$ has a cusp in the spin glass phase. This is due to one more phase transition taking place at that field value, where the magnetization becomes different from zero. In Fig.~\ref{fig:magCurve_XYRF_T0_C03} we show the magnetization modulus $M$ as a function of the external field intensity $H$, computed at the BP fixed point. We observe a phase transition in the magnetization exactly at the point where $\lambda_\text{BP}$ has a cusp. So on the right of the cusp we have an unmagnetised spin glass phase, while on the left we have a mixed phase, i.\,e.~spin glass with non-zero magnetization, usual for disordered systems with a directional bias in the couplings or in the external field. It is worth reminding that the location of this phase transition is only approximate, as one should use an ansatz with RSB in order to compute it exactly \cite{CastellaniEtAl2005}.

\subsection*{Finite-$T$ results and the whole phase diagram}

In order to assess the robustness in temperature of the RSB phase detected with the $T=0$ algorithm, let us move to the finite-$T$ numerics. For large enough values of~$T$, we exploit the \textit{cooling} protocol of Ref.~\cite{LupoRicciTersenghi2018}, to speedup the convergence to the correct BP fixed point. For temperature values too small, e.\,g. below $T/J=0.020$ for the $C=3$ case, we instead prefer to look for the BP fixed point separately for each value of the field modulus, exactly as in the $T=0$ case. However, a single run per temperature and field value is here sufficient.

\begin{figure}[!b]
	\centering
	\includegraphics[width=\columnwidth]{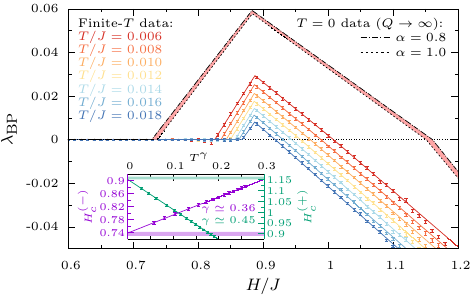}
	\caption{\textit{Main plot}: Stability parameter $\lambda_{\text{BP}}(H)$ for the ferromagnetic XY model in a randomly oriented field of fixed intensity $H$ on the $C=3$ RRG ensemble, for some very small values of the temperature $T$. A piecewise linear fit is also reported, useful to estimate the two critical values of the field at each value of~$T$. The RSB solution survives also at finite temperature, though being not very robust against the temperature increase. For comparison, we finally plot also the $T=0$ data, depicted again as a confidence interval referred to the $Q\to\infty$ extrapolation (while for finite-$T$ data the finite-$Q$ correction is below the precision we are working with). \textit{Inset}: The $T\to 0$ extrapolation of the finite-$T$ critical fields, as in Eq.~(\ref{eq:T_extrap}), is compatible with the confidence interval for the values estimated directly at $T=0$, represented here by the two colored horizontal stripes. Notice the two different y-axis scales, in $J$ units, that refer to the extrapolation of the two different branches of the RS instability line.}
	\label{fig:lambdaBP_XYRF_C03}
\end{figure}

As a first evidence of the reliability of the $T=0$ algorithm, the presence of the RSB region is revealed also by the finite-$T$ results, exactly in the same region where it was expected. However, the robustness in temperature of this RSB phase is very weak, as it can be appreciated in Fig.~\ref{fig:lambdaBP_XYRF_C03}, where we plotted the $\lambda_{\text{BP}}(H)$ curves of the stability parameter for some very small values of the temperature, down to the numerical precision allowed by the finite-temperature algorithm. For comparison, we also reported the data collected directly at $T=0$, as usual represented by the confidence interval $\alpha\in[0.8,1.0]$ for the $Q\to\infty$ extrapolation. At variance, the finite-$T$ data are substantially $Q$-independent already at $Q=64$ within our numerical precision.

Finite-$T$ results for the $C=3$ case are also useful to double-check the reliability of the critical values $H^{(+,-)}_c$ obtained exactly at $T=0$. Indeed, being $Q$-independent (within the error bars represented) at $T>0$, we expect the $T\to 0$ extrapolation of the critical values $H^{(+,-)}_c(T)$ to match the $T=0$ values. It is exactly what can be observed in the inset of Fig.~\ref{fig:lambdaBP_XYRF_C03}, where a square-root-like extrapolation is exploited:
\begin{equation}
	H^{(\pm)}_c(T) \sim H^{(\pm)}_c \mp c^{(\pm)} \, T^{\gamma}
	\label{eq:T_extrap}
\end{equation}
with $\gamma \simeq 0.36$ for the lower branch and $\gamma 	\simeq 0.45$ for the upper branch. The same fit also allows us to easily evaluate the temperature $T^*$ of the tricritical point, that is $T^*/J=0.012(1)$ for $C=3$.

Analogous results hold for the $C=4$ case, where the RSB region can still be detected through the finite-$T$ numerics. Instead, $T^*$ becomes too small for $C \geqslant 5$ to be detected at finite $T$ (last column of Tab.~\ref{tab:Tc_vs_C}).

\begin{figure}[!t]
	\centering
	\includegraphics[width=\columnwidth]{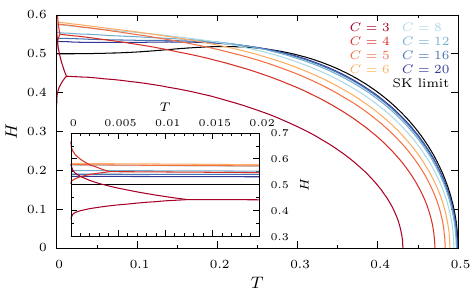}
	\caption{\textit{Main plot:} Phase diagram of the RFXY model with $H_i=H$ for each site and $\phi_i\sim\mathrm{Unif}[0,2\pi)$, for several values of the connectivity $C$ of the RRG ensemble. The strength of the coupling, $J=1/(C-1)$, is such to allow a proper match with the SK fully-connected ($C\to\infty$) case, whose critical line is given by condition~(\ref{eq:XYRF_dense_Mx_stability}). Large values of $C$ only allow for the usual paramagnetic (top right corner) and ferromagnetic (bottom left corner) phases, while for $C=3$ and $C=4$ a small glassy region can be detected close to the zero-temperature axis. \textit{Inset:} Zoom of the glassy region, numerically visible at finite $T$ only for $C=3$ and $C=4$.}
	\label{fig:phase_diag_severalC}
\end{figure}

The entire ($H,T$) phase diagram for the ferromagnetic RFXY model on RRGs can be finally depicted in Fig.~\ref{fig:phase_diag_severalC} for several values of the connectivity $C$ of the graph, again suitably rescaling the couplings by $1/(C-1)$. As an interesting comparison, we also depict the $C\to\infty$ curve, corresponding to the SK limit, that has been analytically obtained in App.~\ref{app:SK_limit} by performing the dense limit of the BP equations, as in Ref.~\cite{LupoRicciTersenghi2018}. The convergence of the large-$C$ curves toward the fully-connected one is quite fast, as already observed in similar models~\cite{LupoRicciTersenghi2018}. Then, an interesting feature can be appreciated already for the $C=16$ and $C=20$ cases, and even more clearly in the SK limit: the non-monotonicity of the critical line between the paramagnetic and the ferromagnetic line.
As discussed in detail in App.~\ref{app:SK_limit}, the curve we are drawing in Fig.~\ref{fig:phase_diag_severalC} is actually the spinodal curve of the paramagnetic phase. In the large $C$ limit it becomes non-monotonic as a consequence of the phase transition becoming first order at low enough temperatures. In App.~\ref{app:SK_limit} we show the exact location of the coexistence phase in the $C=\infty$ limit. For finite $C$ values we have just checked that the saddle point equations admit a unique solution, i.e.\ no coexistence phase, for all $C$ values such that the curve $H_c(T)$ does not change concavity ($C \lesssim 8$).

\section{The RSB phase and the low temperature physics of the RFXY model}
\label{sec:RSB_phase}

\subsection*{Robustness of the RSB phase}

At this point, it is clear that it is the \textit{combined} presence of two key elements that makes possible the onset of the glassy phase in our model: the extreme dilution of interactions (giving a strong spatial heterogeneity) and the continuous nature of the spins. Indeed, as soon as the connectivity $C$ is increased above a certain value, the RSB area can no longer be detected even via the zero-temperature algorithm. Whether its width is still finite for large but finite $C$, or it is identically zero, remains an open question. However, as previously shown, it is no longer appreciable for $C \gtrsim 8$ to the best of our numerical efforts. Analogously, if not performing accurately the $Q\to\infty$ extrapolation, we get an enhanced stability for the RS solution, with values of $\lambda^{(Q)}_{\text{BP}}$ systematically smaller than their $Q\to\infty$ extrapolation.

If our hypothesis were correct, the RSB phase should be robust against small changes in the field distribution, provided there is no strong directional bias that could inhibit the rotational freedom of the spins. In order to check this, let us change our field distribution from the one with fixed modulus $H_i=H$ and random direction $\phi_i$ uniformly drawn from the flat distribution, to the Gaussian one
\begin{equation}
\begin{split}
	&H_{i,x} \sim \mathrm{Gauss}(\mu_x,\sigma^2)\\
	&H_{i,y} \sim \mathrm{Gauss}(0,\sigma^2)
\end{split}
\label{eq:Gauss_field}
\end{equation}
where we decided to orient the potential ferromagnetic bias along the $\hat{x}$ axis ($\mu_y=0$) without any loss of generality. Moreover, let us just focus on the $T=0$ plane, where the effects of RSB (if any) should be more evident and hence more easily detectable.

\begin{figure}[!t]
	\centering
	\includegraphics[width=\columnwidth]{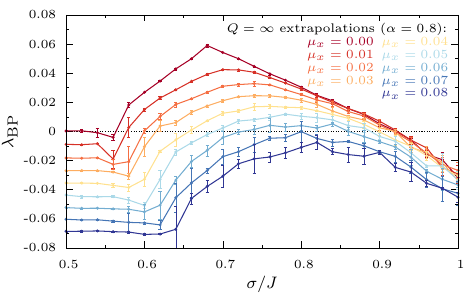}
	\caption{Stability parameter $\lambda_{\text{BP}}$ for the ferromagnetic XY model in a Gaussian-distributed random field as in Eq.~(\ref{eq:Gauss_field}), for the $C=3$ RRG ensemble of graphs at $T=0$. The $Q\to\infty$ curves are again obtained by a power-law extrapolation $Q^{-\alpha}$; only the curves for $\alpha=0.8$ are represented for clarity, with the curves for $\alpha=1.0$ being slightly below. The difference in the location of the critical points according to the different extrapolations can be appreciated in Fig.~\ref{fig:phase_diag_XYGF_T0}. It is evident how the RSB region --- quite wide when the $\mathrm{O}(2)$ symmetry of the field is unbroken --- shrinks to zero when a sufficiently large ferromagnetic bias is inserted. $\mu_x$ values are again measured in $J$ units.}
	\label{fig:lambdaBP_XYGF_C03_T0_Qinfty}
\end{figure}

Also in this case, the infinite-$Q$ extrapolation can be performed by means of a power-law fit, with the exponent $\alpha$ reliably belonging to the same interval $[0.8,1.0]$ as for the randomly-oriented field with constant modulus. Again, in the $\mathrm{O}(2)$-symmetric case ($\mu_x=\mu_y=0$), the zero-temperature algorithm exhibits a RS instability in a non-zero-measure interval of $\sigma$ values (top curve in Fig.~\ref{fig:lambdaBP_XYGF_C03_T0_Qinfty}; for clarity, only the extrapolated curve for $\alpha=0.8$ is shown).

Furthermore, by exploiting the field distribution in Eq.~(\ref{eq:Gauss_field}), we can evaluate the effects on the width of the RSB region when the $\mathrm{O}(2)$ symmetry is explicitly broken, e.\,g. by means of a tiny directional bias $\mu_x$ along the $\hat{x}$ axis, as previously suggested. As soon as $\mu_x$ is switched on, the ferromagnetic phase disappears from the phase diagram, being the $\mathrm{O}(2)$ symmetry already explicitly broken by the external field. However, the zero-temperature algorithm is still able to detect a RS instability, in a region of $\sigma$ values that continuously shrinks to zero with the increase of $\mu_x$ (further curves in Fig.~\ref{fig:lambdaBP_XYGF_C03_T0_Qinfty}). The end-point of this phase is then reached at a certain critical value of the ferromagnetic bias, $\mu_x^*$, above which any $\sigma$ value can only yield a paramagnetic ordering. Of course, the estimate of the value of $\mu^*_x$ depends on the $\alpha$ exponent used for the extrapolation; this systematic error is then taken into account by assigning a suitable uncertainty to the estimate of $\mu^*_x$ (in units of $J$)
\begin{equation}
	\mu^*_x = 0.065 \pm 0.003
\end{equation}

\begin{figure}[!t]
	\centering
	\includegraphics[width=\columnwidth]{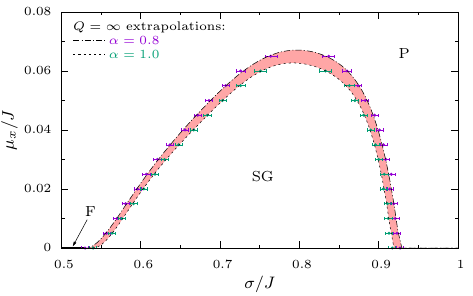}
	\caption{Phase diagram on the $T=0$ plane of the RFXY model with the Gaussian-distributed external field as in Eq.~(\ref{eq:Gauss_field}). The light-red region indicates the confidence interval of the $Q\to\infty$ extrapolated lines, bounded by the fitted datapoints for $\alpha=0.8$ (purple points on the upper curve) and $\alpha=1.0$ (green points on the lower curve). The black curves are given by a 7th-order polynomial fit over the extrapolated datapoints. The glassy RSB region (SG) is the one under such area, while the remaining portion of the $T=0$ plane corresponds to the paramagnetic phase (P). Finally, the ferromagnetic phase (F) is restricted to the $\mu_x=0$ axis for $\sigma < \sigma^{(-)}_c$, since suddenly vanishing when switching on $\mu_x$.}
	\label{fig:phase_diag_XYGF_T0}
\end{figure}

From a further inspection of Fig.~\ref{fig:lambdaBP_XYGF_C03_T0_Qinfty}, we can realize that there are two main effects caused by the switch-on of the ferromagnetic bias $\mu_x$. First of all, a smoothing of the $\lambda_{\text{BP}}(\sigma)$ curves around the point of maximal instability, directly caused by the breaking of the $\mathrm{O}(2)$ symmetry of the field, and the consequent inhibition of the rotational freedom of the spins in the configuration space. The disappearance of the cusp for positive values of $\mu_x$ hence reflects the corresponding disappearance of the phase transition (still present for $\mu_x=0$) between the mixed and the unmagnetized glassy phase, being left only with a magnetized spin glass phase. Secondly, a south-east shift of the curves, resulting in a larger $\mu_x$-dependence of the critical point $\sigma^{(-)}_c$ compared to $\sigma^{(+)}_c$.

Finally, the shift of the $\lambda_{\text{BP}}(\sigma)$ curves has a further important consequence, that we will further discuss in the following section: the sudden loss of the marginal ferromagnetic phase as soon as the directional bias $\mu_x$ is switched on.

The resulting phase diagram $(\mu_x,\sigma)$ on the $T=0$ plane is finally shown in Fig.~\ref{fig:phase_diag_XYGF_T0}, where the confidence interval for the $Q\to\infty$ extrapolation of the critical lines is again represented by the light-red region. If we plotted the whole $3d$ phase diagram by also taking into account the temperature, we would also in this case recognize a shrinking to zero of the RSB region on the $\mu_x=0$ surface when increasing the temperature, so resulting in the overall picture of a tiny RSB region, weakly robust with respect to the increase of either the temperature or the field anisotropy. At variance, the ferromagnetic phase would remain relegated in the $\mu_x=0$ plane.

Concluding, we verified the robustness of the RSB region with respect to slight changes in the field distribution, confirming the strong sparsity and the rotational freedom of the spins as the major actors of the onset of the glassiness, while ruling out unessential details of the field distribution as a small ferromagnetic bias (explicitly breaking the rotational symmetry) or the randomness in the local field strength.

\subsection*{Marginality of the ferromagnetic phase}

Another surprising feature of the random field XY model can be found in the ferromagnetic phase, just below the glassy phase. Indeed, as already suggested by the analysis of Fig.~\ref{fig:lambdaBP_XYRF_C03}, the ferromagnetic phase at $T=0$ is characterized by a marginal stability, that is preserved even when switching on the temperature.

The connection with the continuous nature of the XY spins is straightforward: as long as $Q$ is finite, then $\lambda_{\text{BP}}$ is strictly negative (although very close to zero) for $H<H^{(-)}_c$ and the corresponding ferromagnetic solution is strictly stable. Conversely, when properly extrapolating to the continuous limit ($Q\to\infty$), we get a value for the stability parameter compatible with zero in the whole ferromagnetic phase.

In general the presence of an external field --- even better if disordered and site-uncorrelated ---  eliminates any potential zero mode due to symmetries of the Hamiltonian. In this case, notwithstanding the presence of the random field, we observe a marginal ferromagnetic phase. A careful analysis of this marginally stable solution reveals that, although the global magnetization is non-zero (due to the ferromagnetic long-range order), all XY spin variables can coherently rotate without paying a substantial energy cost thanks to the global $\mathrm{O}(2)$ symmetry of the random field distribution. In other words, although on each site the random field locally breaks the $\mathrm{O}(2)$ symmetry and introduces a preferential direction, at the global level the field distribution is $\mathrm{O}(2)$ symmetric and thus has a zero mean: this allows the global magnetization to rotate without paying a substantial energy cost.

In practice for finite $N$ systems, infinitesimal perturbations can propagate through the entire system, thus leading to a slow global rotation with a global energetic cost that goes to zero in the thermodynamic limit. This has been observed also while running the PDA at finite temperatures for long enough times.

As an evidence of what explained above, when explicitly breaking the symmetry of the external field --- e.\,g. by switching on $\mu_x$ in the Gaussian case --- the marginality is suddenly lost, resulting in a negative value for $\lambda_{\text{BP}}$, which is smaller the larger $\mu_x$ (see Fig.~\ref{fig:lambdaBP_XYGF_C03_T0_Qinfty}).

Notice finally that the marginality of the whole ferromagnetic phase is entirely related to the Goldstone modes of the model that survive notwithstanding the random field, thanks to its unbroken $\mathrm{O}(2)$ symmetry. A more physical interpretation of these zero modes in terms of correlations will be then provided in Sec.~\ref{sec:corr_funcs}. At variance with this, there are other continuous ferromagnetic models that also exhibit marginality in the ordered phase~\cite{CavagnaEtAl2018}. However, its nature is very different and related to diverging fluctuations in the modulus of the spin vectors, which is instead fixed in the XY model we are studying.

\subsection*{Meaning of the RS instability}

At this point, we go back to the argument provided in Ref.~\cite{KrzakalaEtAl2010} against the presence of a RSB phase in the RFIM. The basic observation is that connected correlation functions $\braket{\sigma_i\,\sigma_j}_c=\braket{\delta\sigma_i\,\delta\sigma_j}$ are always nonnegative if the model is purely ferromagnetic, due to the solely possibility of longitudinal perturbations of equilibrium configurations for Ising spins. So ferromagnetic susceptibility $\chi_{\text{F}}$ represents an upper bound for the spin glass one $\chi_{\text{SG}}$ and hence no transition toward a spin glass phase can be realized out of the zero-measured $P-F$ critical point.

In fact, vector spins allow not only longitudinal perturbations, but also transverse ones around the equilibrium configuration, as thoroughly studied for the spin glass XY model in an external field~\cite{LupoRicciTersenghi2018}. More formally, connected correlations for $m$-dimensional vector spins are no longer scalars, but $m\times m$ matrices $\braket{\delta\sigma_{i,\mu}\,\delta\sigma_{j,\nu}}$ --- with $m=2$ for the XY model --- whose entries can be either positive or negative, irrespective of the presence of the ferromagnetic couplings, and being related to the longitudinal and the transverse responses to excitations in the system. Finally, this is enough to provide a negative effective correlation even between nearest-neighbor spins joined by a positive coupling, and eventually to cause enough frustration to make room for glassiness. To check this, let us compute connected correlations in the next Section.

\section{Correlation function in the diluted XY model}
\label{sec:corr_funcs}

In order to compute the matrix $\mathcal{M}$ of connected correlations of two spins $\boldsymbol{\sigma}_i$ and $\boldsymbol{\sigma}_j$
\begin{equation}
\begin{split}
	\mathcal{M}(i,j) &\equiv \braket{\boldsymbol{\sigma}_i\,\boldsymbol{\sigma}^{\intercal}_j}_c \\
	&= \int d\boldsymbol{\sigma}_i \, d\boldsymbol{\sigma}_j \, \mathbb{P}(\boldsymbol{\sigma}_i,\boldsymbol{\sigma}_j) \, \boldsymbol{\sigma}_i \, \boldsymbol{\sigma}^{\intercal}_j\\
	&\qquad -\int d\boldsymbol{\sigma}_i \, \mathbb{P}(\boldsymbol{\sigma}_i) \, \boldsymbol{\sigma}_i\int d\boldsymbol{\sigma}_j \, \mathbb{P}(\boldsymbol{\sigma}_j) \, \boldsymbol{\sigma}^{\intercal}_j\\
	&= \int d\boldsymbol{\sigma}_i \, d\boldsymbol{\sigma}_j \, \mathbb{P}_c(\boldsymbol{\sigma}_i,\boldsymbol{\sigma}_j) \, \boldsymbol{\sigma}_i \, \boldsymbol{\sigma}^{\intercal}_j
		\label{eq:connected_corr_mag}
\end{split}
\end{equation}
we need to know their connected joint probability distribution, defined as:
\begin{equation}
	\mathbb{P}_c(\boldsymbol{\sigma}_i,\boldsymbol{\sigma}_j) \equiv \mathbb{P}(\boldsymbol{\sigma}_i,\boldsymbol{\sigma}_j) - \mathbb{P}(\boldsymbol{\sigma}_i) \, \mathbb{P}(\boldsymbol{\sigma}_j)
	\label{eq:connected_joint_P}
\end{equation}

Let us focus on the $m=2$ case, so moving again to angular variables $\theta$'s. When nodes $i$ and $j$ are at distance $r=1$, BP already provides the probability distributions we need. Indeed, from Eqs.~(\ref{eq:BP_XY_onePoint}) and~(\ref{eq:BP_XY_twoPoints}):
\begin{equation}
	\mathbb{P}_c(\theta_i,\theta_j) = \eta_{ij}(\theta_i,\theta_j) - \eta_i(\theta_i)\,\eta_j(\theta_j)
\end{equation}
and in turn these ones can be computed by e.\,g. directly sampling from the fixed-point cavity distribution $\mathbb{P}^*[\eta_{i\to j}]$ in the PDA. However, the most relevant information content is hidden into the long-distance decay of correlations, that have to be computed in a different way.

\subsection*{The computation along a chain}

To this aim, we will exploit the well known computation of correlation functions along a chain~\cite{MoroneEtAl2014}, generalizing it from the Ising case to the XY case. Let us start from a $r=1$-long chain, where the full joint probability distribution reads, from Eq.~(\ref{eq:BP_XY_twoPoints})
\begin{equation}
	\mathbb{P}(\theta_i,\theta_j) \cong e^{\,\beta J\cos{(\theta_i-\theta_j)}}\,\eta_{i\to j}(\theta_i)\,\eta_{j\to i}(\theta_j)
\end{equation}
neglecting the obvious normalization. We can distinguish two key ingredients: \textit{i)} the ``external legs'' $\eta_{i\to j}(\theta_i)$ and $\eta_{j\to i}(\theta_j)$, and \textit{ii)} the ``amputated correlation'' $\mathbb{P}_a(\theta_i,\theta_j) = \exp{\{\beta J\cos{(\theta_i-\theta_j)}\}}$. Then, the connected joint probability distribution can be computed as in Eq.~(\ref{eq:connected_joint_P}), getting the single-variable probability distributions by marginalization.

When moving to a $r=2$-long chain, by e.\,g. extending it on the right side, we have to add the contribution of the external field $\boldsymbol{H}_j$ acting on site $j$, the link $\exp{\{\beta J\cos{(\theta_j-\theta_k)}\}}$ with the new ``right leg'' ending in site $k$, and the remaining $C-2$ contributions to $j$ from the sides of the chain, namely the neighbors of site~$j$ different from $i$ and $k$. Finally, the new amputated contribution is obtained by integrating over $\theta_j$:
\begin{equation}
\begin{split}
	\mathbb{P}_a(\theta_i,\theta_k) &\cong \int d\theta_j \, \mathbb{P}_a(\theta_i,\theta_j) \, e^{\,\beta J\cos{(\theta_j-\theta_k)}}\\
	 &\qquad \times e^{\,\beta H\cos{(\theta_j-\phi_j)}} \, \prod_{l=1}^{C-2}\hat{\eta}_{l\to j}(\theta_j)
\end{split}
\end{equation}
where $\hat{\eta}_{l\to j}(\theta_j)$ is the convolution of the cavity message $\eta_{l\to j}(\theta_l)$ with the compatibility function of the interaction between sites $l$ and $j$:
\begin{equation}
	\hat{\eta}_{l\to j}(\theta_j) \cong \int d\theta_l \, \eta_{l\to j}(\theta_l) \, e^{\,\beta J\cos{(\theta_l-\theta_j)}}
\end{equation}
The full joint $\mathbb{P}(\theta_i,\theta_k)$ can be then obtained by multiplying the amputated one with the external leg contributions, $\eta_{i\to j}(\theta_i)$ and $\eta_{k\to j}(\theta_k)$, and in turn the connected one from it as before.

In this way, we can actually iterate the procedure and so obtain the connected joint probability distribution $\mathbb{P}_c(\boldsymbol{\sigma}_i,\boldsymbol{\sigma}_j)$ of two generic spins $\boldsymbol{\sigma}_i$ and $\boldsymbol{\sigma}_j$ at distance~$r$ along a chain.

From the numerical point, being one-variable probability distributions approximated as $Q$-component vectors, then two-variable probability distributions are represented as $Q \times Q$ matrices. Finally, these matrices can be projected onto the $m \times m$ ones in the magnetization space as in Eq.~(\ref{eq:connected_corr_mag}), which for the XY model explicitly becomes
\begin{equation}
	\left\{
	\begin{aligned}
		&\mathcal{M}_{xx}(i,j) \equiv \braket{\sigma_{i,x}\,\sigma_{j,x}}_c\\
			&\qquad\quad = \frac{1}{(2\pi)^2}\int d\theta_i \, d\theta_j \, \mathbb{P}_c(\theta_i,\theta_j) \, \cos{\theta_i} \, \cos{\theta_j}\\
		&\mathcal{M}_{xy}(i,j) \equiv \braket{\sigma_{i,x}\,\sigma_{j,y}}_c\\
			&\qquad\quad = \frac{1}{(2\pi)^2}\int d\theta_i \, d\theta_j \, \mathbb{P}_c(\theta_i,\theta_j) \, \cos{\theta_i} \, \sin{\theta_j}\\
		&\mathcal{M}_{yx}(i,j) \equiv \braket{\sigma_{i,y}\,\sigma_{j,x}}_c\\
			&\qquad\quad = \frac{1}{(2\pi)^2}\int d\theta_i \, d\theta_j \, \mathbb{P}_c(\theta_i,\theta_j) \, \sin{\theta_i} \, \cos{\theta_j}\\
		&\mathcal{M}_{yy}(i,j) \equiv \braket{\sigma_{i,y}\,\sigma_{j,y}}_c\\
			&\qquad\quad = \frac{1}{(2\pi)^2}\int d\theta_i \, d\theta_j \, \mathbb{P}_c(\theta_i,\theta_j) \, \sin{\theta_i} \, \sin{\theta_j}
	\end{aligned}
	\right.
	\label{eq:connected_corr_mag_XY}
\end{equation}
with $i$ and $j$ being at distance $r$ along the given chain. These $2 \times 2$ matrices encode the longitudinal and transverse responses to small excitations in the system, which we believe are the ones explicitly bearing the signatures of the marginality of the ferromagnetic phase and of the RS instability of the glassy phase.

\begin{figure*}[!t]
	\centering
	\includegraphics[width=\textwidth]{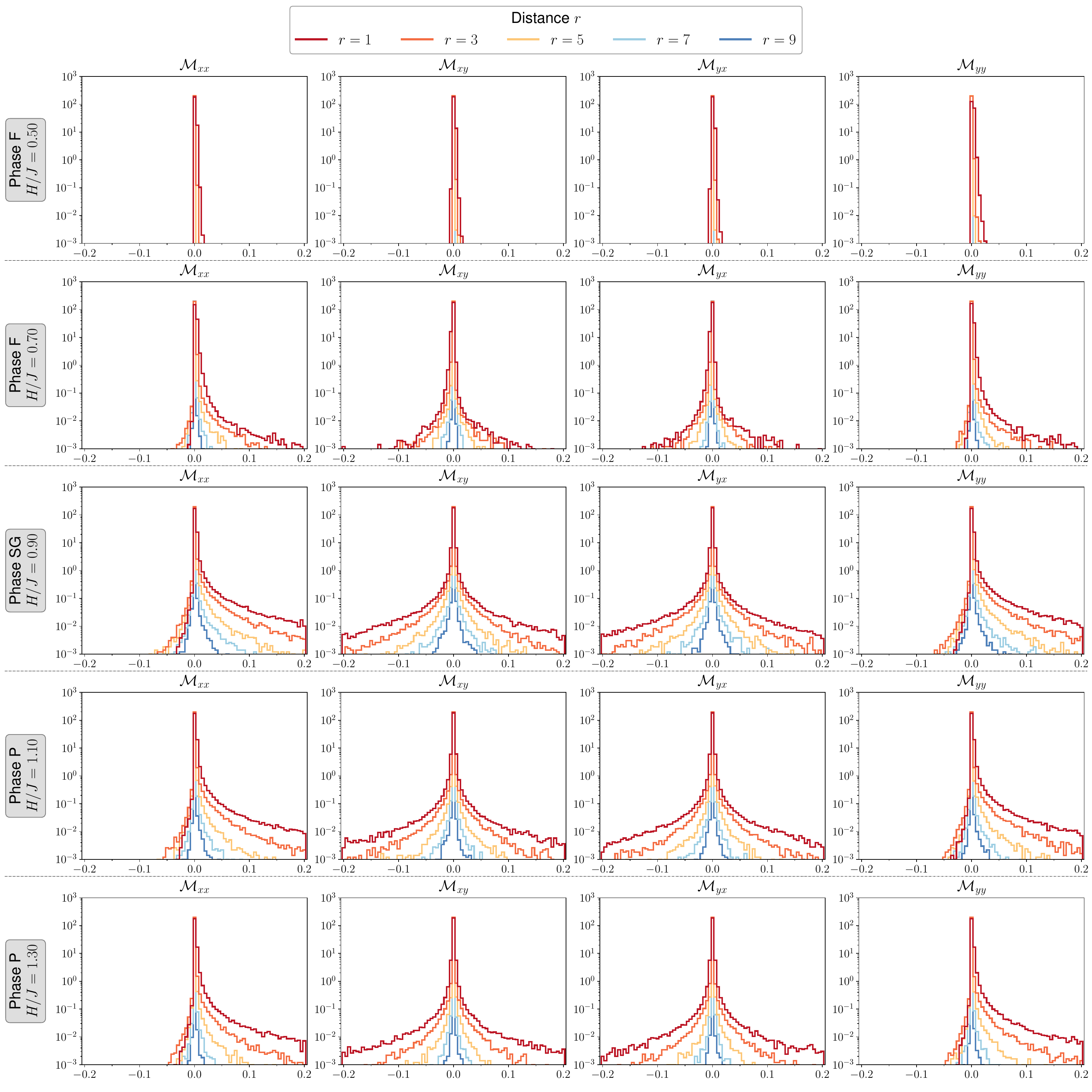}
	\caption{Probability distributions of the entries $\mathcal{M}_{\mu\nu}(r=|i-j|)$ at different distances $r$ over $10^6$ realizations of a chain, computed according Eqs.~(\ref{eq:connected_corr_mag_XY}) in the $C=3$ case. We set $T/J=0.01$ and five different values of $H$ so to study all the three phases: ferromagnetic, spin glass and paramagnetic. The corresponding critical values of the field modulus are $H^{(-)}_c=0.843(2)$ for the $F-SG$ transition and $H^{(+)}_c=0.973(2)$ for the $SG-P$ transition, in units of $J$. The two axes keep the same ranges for all the panels, to favour the comparison; notice also the log scale on the vertical axis, useful to appreciate the different decay at large distances. The different behavior of the distributions in the three phases is discussed in the main text.}
	\label{fig:corrHisto_allPhases_T0_01_5fields}
\end{figure*}

\begin{figure*}[!t]
	\centering
	\includegraphics[width=\textwidth]{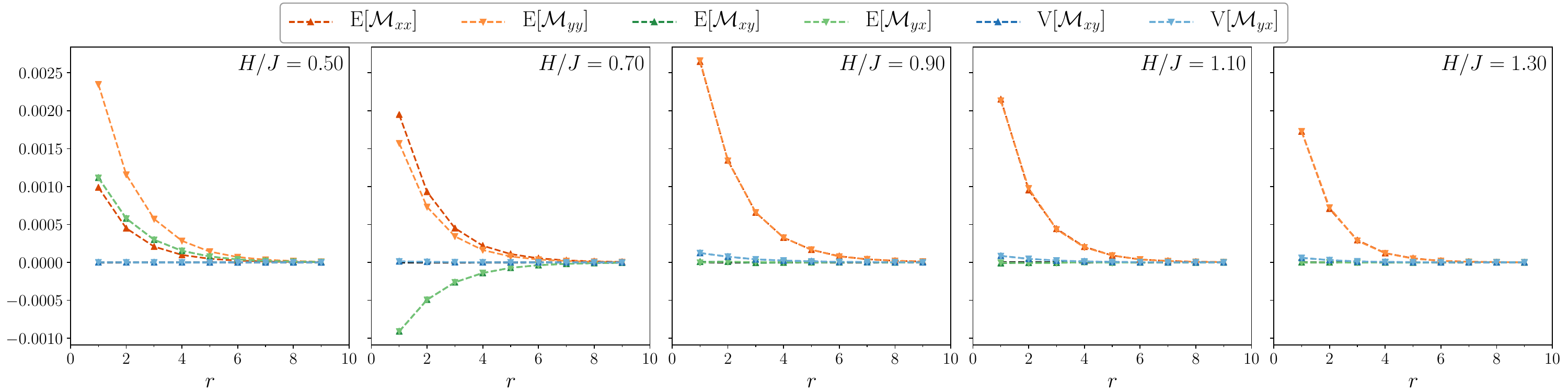}
	\caption{Mean $\mathbb{E}[\cdot]$ and variance $\mathbb{V}[\cdot]$ of the entries $\mathcal{M}_{\mu\nu}(r)$ when increasing the distance $r$, from the same $10^6$ realizations of a chain used for Fig.~\ref{fig:corrHisto_allPhases_T0_01_5fields}. Being $C=3$ and having set $T/J=0.01$, we have a ferromagnetic phase for $H/J=\{0.50,0.70\}$, a spin glass phase for $H/J=0.90$ and a paramagnetic phase for $H/J=\{1.10,1.30\}$. Vertical axes have all the same range, with values reported in the leftmost panel. The different behavior of the moments in the three phases is discussed in the main text.}
	\label{fig:corrDecay_allPhases_T0_01}
\end{figure*}

\subsection*{Decay of correlations}

In what we did before, the RS instability has been detected by looking at growth rate $\lambda_{\text{BP}}$ (in the time domain) of the global norm of the perturbations to the BP fixed point $\mathbb{P}^*[\eta_{i\to j}]$ --- both at finite and zero temperature ---, which of course represents a reliable tool, providing results that match with the analytic predictions where the latter ones are available. However, once having in our hands connected correlation functions, we can study their exponential decay with the distance
\begin{equation}
		\mathcal{C}(r) \sim e^{-r/\xi}
\end{equation}
and hopefully match the divergence of their correlation length $\xi$ with the critical point already detected before.

In order to do that, since correlations are computed along a chain --- and eventually averaged over a large enough number of chain realizations ---, the corresponding correlations on the original graph ensemble can be recovered by multiplying $\mathcal{C}(r)$ by the inverse branching ratio to the $r$-th power, $(C-1)^r$, i.\,e. the average number of neighbors at distance $r$ on the graph. The resulting correlation length $\xi'$ is finally expected to diverge exactly in correspondence of second-order transitions detected by the condition $\lambda_{\text{BP}}=0$ in the PDA.

A first, rapid check of the coherence of the two approaches can be attained by looking at the average decay rate of the $\ell_2$ norm of the $Q \times Q$ matrix representing $\mathbb{P}_c(\theta_i,\theta_j)$. For example, if we fix the temperature $T$ and perform an annealing in the field modulus $H$, then compute the decay rate $1/\xi'$ at each BP fixed point, we exactly recover the critical values $H^{(-)}_c(T)$ and $H^{(+)}_c(T)$ given by the well assessed stability study in the PDA.

\begin{figure*}[!t]
	\centering
	\includegraphics[width=\textwidth]{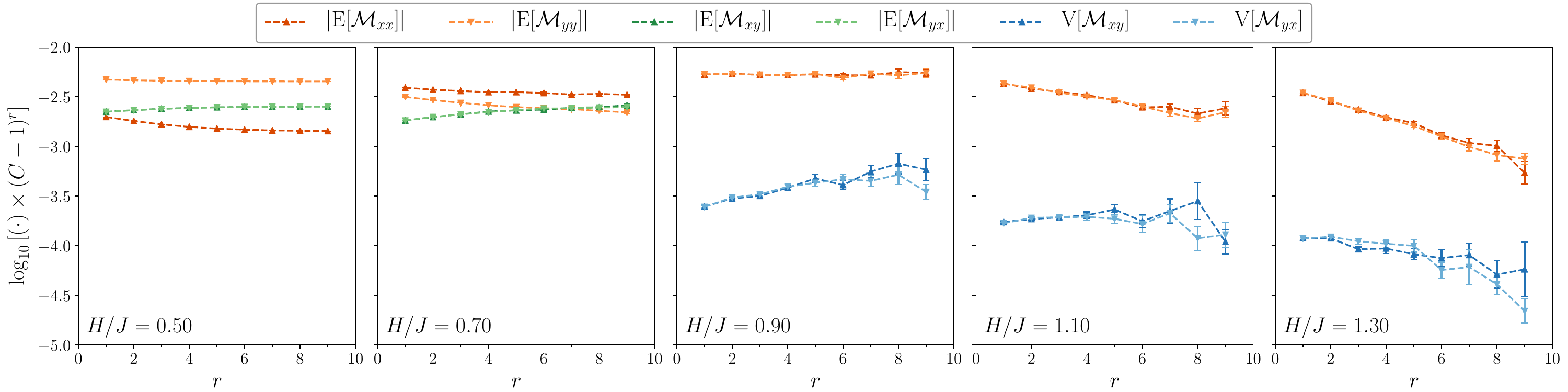}
	\caption{Same mean $\mathbb{E}[\cdot]$ and variance $\mathbb{V}[\cdot]$ of the entries $\mathcal{M}_{\mu\nu}(r)$ as in Fig.~\ref{fig:corrDecay_allPhases_T0_01}, now corrected with the inverse branching ratio $(C-1)^r$ so to move from the chain ensemble to the $C$-RRG graph ensemble. Vertical axes have all the same range, with values in the $\log_{10}$ scale reported in the leftmost panel. It is very clear the marginality of the ferromagnetic phase, the instability of the glassy phase and the stability of the paramagnetic phase. When $r \gtrsim 10$, finite-size fluctuations become too relevant and overcome the expected large-$r$ behavior.}
	\label{fig:corrDecayRescaled_allPhases_T0_01}
\end{figure*}

\subsection*{Critical correlations}

However, the knowledge of correlation functions gives us by far more information than the location of the critical points. Indeed, we can characterize which correlations are the relevant ones in each thermodynamic phase, and then also understand which of them become critical at each phase transition.

To this aim, let us look at the single entries of matrix $\mathcal{M}(r=|i-j|)$. Being the heterogeneity a key feature of disordered systems on diluted topologies --- even more in our case, where it is a necessary ingredient for the onset of the RSB phase ---, we firstly look at their probability distribution over $10^6$ realizations of different chains, every time picking the necessary cavity messages from the BP fixed point $\mathbb{P}^*[\eta_{i\to j}]$. In Fig.~\ref{fig:corrHisto_allPhases_T0_01_5fields} we reported them for five different values of the field modulus, for the $C=3$ case at fixed temperature $T/J=0.01$: two in the ferromagnetic phase, one in the middle of the glassy region, and two in the paramagnetic phase.

The general behavior when looking at large distances is a shrinking in the distributions, quantifiable through the decay of both mean and variance with $r$; indeed, we will look at it after. Before, let us qualitatively appreciate the difference between the three phases, in order to validate (or not) our picture. Notice the vertical axis being in the log scale and with the same range in all the panels, in order to better appreciate the different large-$r$ decay of correlations, otherwise not visible if using the linear scale.

In the ferromagnetic phase (first and second rows of panels in Fig.~\ref{fig:corrHisto_allPhases_T0_01_5fields}), a global magnetization of order one characterizes the system. Hence, fluctuations have to take it into account, resulting in histograms of the diagonal entries $\mathcal{M}_{xx}$ and $\mathcal{M}_{yy}$ (namely, the longitudinal responses with respect to the perturbation) almost entirely lying in the positive region. However, at variance with the Ising case, a small tail is also present in the negative region, as a signature of the negative effective correlations that can take place in the XY model, already at distance $r=1$. Then, the two first moment of diagonal entries are in general different, depending on the angle between the direction of the spontaneous breaking of the $\mathrm{O}(2)$ symmetry and the $\hat{x}$ axis (notice indeed the difference between $H/J=0.50$ and $H/J=0.70$, being independent runs and hence having a different direction for the global magnetization). At the same time, the two off-diagonal entries (namely, the transverse responses with respect to the perturbation) acquire a very similar distribution, with the same width and centered around the same mean, which in turn can be either negative (e.\,g. for the run at $H/J=0.70$) or positive (e.\,g. for the run at $H/J=0.50$) depending again on the direction of the global magnetization.

When moving to the paramagnetic phase (fourth and fifth rows of panels in Fig.~\ref{fig:corrHisto_allPhases_T0_01_5fields}), the $\mathrm{O}(2)$ symmetry is no longer spontaneously broken, so that diagonal correlations $\mathcal{M}_{xx}$ and $\mathcal{M}_{yy}$ now have the same distribution, with a fatter tail in the positive domain. Also off-diagonal entries have the same distribution, but now centered around zero and symmetric, due to symmetry arguments.

The spin glass phase (center panels in Fig.~\ref{fig:corrHisto_allPhases_T0_01_5fields}) is quite indistinguishable from the paramagnetic one, according to $\mathcal{M}_{\mu\nu}$ distributions. Indeed, the $\mathrm{O}(2)$ symmetry is locally broken, though along incoherent directions throughout the system, so that the overall picture is quite the same of the paramagnetic one. Diagonal entries behave the same and have a positive mean, while off-diagonal entries have again a symmetric distribution around the zero.

From the qualitative analysis of the features of the distributions in Fig.~\ref{fig:corrHisto_allPhases_T0_01_5fields}, we can now properly choose the moments of such distributions whose decay may become critical at the critical points. Indeed, in the ferromagnetic phase, both the longitudinal and the transverse entries of the matrix $\mathcal{M}$ have a non-vanishing first moment $\mathbb{E}[\mathcal{M}_{\mu\nu}]$, whose decay with the distance $r$ can be well appreciated in the first two panels of Fig.~\ref{fig:corrDecay_allPhases_T0_01}. As already said, the longitudinal ones are always positive, but in general with different values; instead, the transverse ones have the same value, either negative or positive depending on the direction of the global magnetization. At variance, both in the spin glass phase (central panel of Fig.~\ref{fig:corrDecay_allPhases_T0_01}) and in the paramagnetic phase (last two panels of Fig.~\ref{fig:corrDecay_allPhases_T0_01}), we can actually verify that $\mathbb{E}[\mathcal{M}_{xx}]$ and $\mathbb{E}[\mathcal{M}_{yy}]$ lie on the same curve as functions of $r$, while $\mathbb{E}[\mathcal{M}_{xy}]$ and $\mathbb{E}[\mathcal{M}_{yx}]$ are zero within uncertainty bars. So the decay of the transverse components is in fact described by their second moments, respectively $\mathbb{V}[\mathcal{M}_{xy}]$ and $\mathbb{V}[\mathcal{M}_{yx}]$.

To finally check which of these moments become critical at the two transitions, we have to multiply them by the inverse branching ratio $(C-1)^r$ and then evaluate their new slope $1/\xi'$ in the log scale (Fig.~\ref{fig:corrDecayRescaled_allPhases_T0_01}). As expected, the four first moments $\mathbb{E}[\mathcal{M}_{\mu\nu}]$ --- once rescaled with the inverse branching ratio --- cease to exponentially decay with the distance and stay critical in the whole ferromagnetic phase. Of course, it is not surprising to see that not only transverse fluctuations are critical, but also the longitudinal ones, being related to the well known simultaneous divergence of the longitudinal susceptibility $\chi_{\parallel}$ and the transverse susceptibility $\chi_{\perp}$ in the explicitly unbroken $\mathrm{O}(2)$ symmetry~\cite{Book_PatashinskiiPokrovskii1979}.

The spin glass phase, then, seems to be characterized by a critical decay of the longitudinal first moments (or maybe a slight divergence, within our resolution), while the transverse second moments are clearly showing unstable fluctuations. So the instability of the RS solution in the glassy region seems to be mainly due to long-range transverse fluctuations, as we guessed at the beginning of our analysis.

Finally, the paramagnetic phase is again characterized by integrable fluctuations of both the mean of longitudinal components and the variance of transverse ones (up to $r \simeq 10$, where finite-size fluctuations become too strong), as expected for a RS-stable phase.

\section{Conclusions}
\label{sec:conclusions}

We have obtained the analytic solution to the random field XY model on a RRG at the RS level by solving the BP equations via a discrete approximation and finally taking the continuous limit. We have derived all possible phase diagrams with different field distributions and we have found some surprising results.

The model possesses a RSB phase where spin glass long-range order develops. This phase is present at very low temperatures and is clearly visible only for RRGs of small degree. So a strong sparsity seems to be a key ingredient for the existence of long-range spin glass order, together with the continuous nature of the XY variables (indeed in random field Ising models the RSB phase can not exist). We have also checked that the RSB phase is robust with respect to the field distribution: e.\,g. it survives also in presence of a random field with a non-zero mean, as long as this mean value is small enough.

Another unexpected result is the observation that the ferromagnetic phase is marginally stable. This comes as a surprise, since the presence of a random field locally breaks the $\mathrm{O}(2)$ symmetry and defines a preferential direction for each XY spin. However, if the distribution of the random field is $\mathrm{O}(2)$ symmetric, then the global $\mathrm{O}(2)$ symmetry of the field allows for the coherent rotation of all XY spins without paying an extensive energy cost. These collective fluctuations make the ferromagnetic phase marginally stable.

We have finally identified the correlations that seem to be the most informative in order to identify the kind of long-range order that takes place in the system. Having measured the $2 \times 2$ matrix of correlations between the components of two spins at a fixed distance, we have studied how the distributions of the diagonal and off-diagonal elements evolve with the distance in the different phases. The symmetry of the distribution of the off-diagonal elements is related to the phase symmetry, thus allowing to identify the ferromagnetic phase. Instead the transition from paramagnetic to spin glass phase is signaled by the critical decay of the mean of diagonal elements or the variance of the off-diagonal ones (whose mean is always zero in these phases).

\begin{acknowledgments}
	Most part of this work has been done while C. Lupo was a post-doctoral fellow at the Physics Department of Sapienza University of Rome. This research has been supported by the Simons Foundation (grant No. 454949, G. Parisi) and by the European Research Council (ERC) under the European Unions Horizon 2020 research and innovation programme (grant No. 694925, G. Parisi).
\end{acknowledgments}

\appendix

\section{The SK limit}
\label{app:SK_limit}

The critical line of the random field XY model in the dense (or SK) limit can be easily obtained by performing the $C\to\infty$ limit of the BP equations. Once assessed the general strategy, as thoroughly explained in Ref.~\cite{LupoRicciTersenghi2018}, the random field XY model with ferromagnetic couplings turns out to be a very simple case.

Indeed, given that all the $J$'s take on the same value and are of order $1/(C-1) \sim 1/N$, in the expansion of Eqs.~(A54) and~(A56) in Ref.~\cite{LupoRicciTersenghi2018}, we can just retain the field term and the first-order term in $J$:
\begin{equation}
	h_{i\to j}(\boldsymbol{\sigma}_i) \simeq \boldsymbol{H}_i \cdot \boldsymbol{\sigma}_i + J\sum_{k\in\partial i\setminus j} \boldsymbol{\sigma}_i \cdot \braket{\boldsymbol{\sigma}_k}_k
\end{equation}
Hence, when exploiting the vectorial ansatz for the cavity field, $h_{i\to j}(\boldsymbol{\sigma}_i) \equiv \boldsymbol{\xi}_{i\to j}\cdot\boldsymbol{\sigma}_i$, we get:
\begin{equation}
	\boldsymbol{\xi}_{i\to j} = \boldsymbol{H}_i + J\sum_{k\in\partial i\setminus j} \braket{\boldsymbol{\sigma}_k}_k
	\label{eq:XYRF_dense_xi}
\end{equation}
where the average $\braket{\boldsymbol{\sigma}_k}_k$ can be analytically computed by means of Bessel functions:
\begin{equation}
\begin{split}
	\braket{\boldsymbol{\sigma}_k}_k &\equiv \frac{\int d\boldsymbol{\sigma}_k \exp{[\beta h_{k\to i}(\boldsymbol{\sigma}_k)]}\boldsymbol{\sigma}_k}{\int d\boldsymbol{\sigma}_k \exp{[\beta h_{k\to i}(\boldsymbol{\sigma}_k)]}}\\
	&=\frac{I_1(\beta\xi_{k\to i})}{I_0(\beta\xi_{k\to i})}\,\frac{\boldsymbol{\xi}_{k\to i}}{\xi_{k\to i}}
\end{split}
\end{equation}

When summing over the $\mathcal{O}(N)$ neighbors, the second term in the right hand side of Eq.~(\ref{eq:XYRF_dense_xi}) concentrates around its mean, namely the vector of the global magnetization $\boldsymbol{M}$. Dropping the edge notation in the dense limit, vectors $\boldsymbol{\xi}$'s are hence random variables satisfying
\begin{equation}
\begin{aligned}
	\xi_x &= H\cos{\phi}+M_x\\
	\xi_y &= H\sin{\phi}+M_y
\end{aligned}
\end{equation}
where the randomness is just given by the field direction~$\phi$.

Explicitly breaking the symmetry along the $\hat{x}$ axis, we have that $M_y$ identically vanishes in both the paramagnetic and the ferromagnetic phase. We are only left with the self-consistency equation for $M_x$:
\begin{equation}
	M_x = \mathbb{E}_{\phi}\bigl[\xi_x\bigr] = \mathbb{E}_{\phi}\biggl[\frac{I_1(\beta\xi)}{I_0(\beta\xi)}\,\frac{\xi_x}{\xi}\biggr]
	\label{eq:XYRF_dense_Mx}
\end{equation}
where $\xi_x=H\cos{\phi}+M_x$, $\xi_y=H\sin{\phi}$ and $\xi^2=\xi_x^2+\xi_y^2$.

\begin{figure}
    \centering
    \includegraphics[width=\columnwidth]{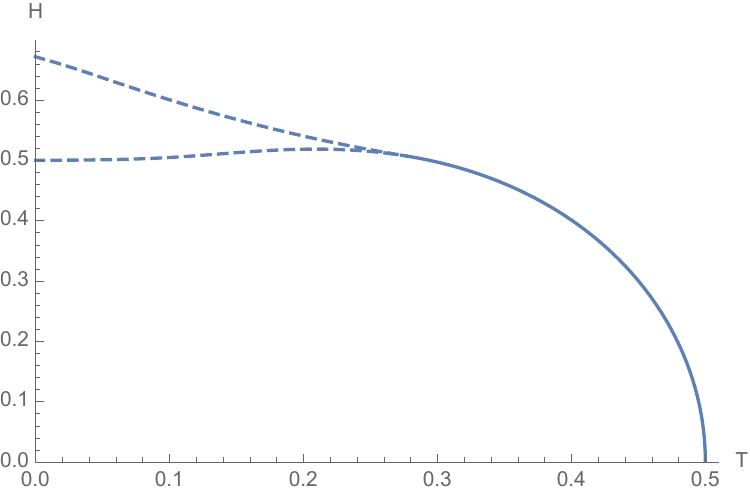}
    \caption{The phase diagram of the random field XY model in the $C=\infty$ limit (fully connected graph). Along the full curve the phase transition to the ferromagnetic phase is continuous, while dashed curves are the spinodal lines for the paramagnetic (lower) and ferromagnetic (upper) phases.}
    \label{fig:Cinfty}
\end{figure}

Eq.~(\ref{eq:XYRF_dense_Mx}) admits always the trivial solution $M_x=0$. The non-trivial solution may appear continuously or discontinuously. For a continuous phase transition the critical line can be computed by expanding Eq.~(\ref{eq:XYRF_dense_Mx}) to linear order in $M_x$
\begin{equation}
\begin{aligned}
	M_x &= \mathbb{E}_{\phi}\biggl[\frac{I_1(\beta\xi)}{I_0(\beta\xi)}\,\frac{\xi_x}{\xi}\biggr]\\
		&\simeq \mathbb{E}_{\phi}\Biggl[\frac{d}{d\,M_x}\biggl(\frac{I_1(\beta\xi)}{I_0(\beta\xi)}\,\frac{\xi_x}{\xi}\biggr)\biggr{|}_{M_x=0}\Biggr] M_x\\
		&=\frac{\beta}{2}\biggl(1-\frac{I^2_1(\beta H)}{I^2_0(\beta H)}\biggr)M_x\;,
\end{aligned}
\end{equation}
providing the condition
\begin{equation}
	\frac{\beta}{2}\biggl(1-\frac{I^2_1(\beta H)}{I^2_0(\beta H)}\biggr)=1\;.
	\label{eq:XYRF_dense_Mx_stability}
\end{equation}
The presence of a first order transition is signalled by a positive third order coefficient (the second order one is always null) and this happens for temperatures below the tricritical point $(T_{tc},H_{tc})=(0.276678,0.507095)$.
For $T<T_{tc}$ one needs to solve Eq.~(\ref{eq:XYRF_dense_Mx}) without any approximation and finds a coexistence region where the equation admits 2 solutions. In Fig.~\ref{fig:Cinfty} the coexistence region is the one between the two dashed curves that represent respectively the spinodal lines for the paramagnetic phase (lower curve) and for the  ferromagnetic phase (upper curve). At $T=0$ these spinodal lines ends in $H=0.5$ and $H=0.671514$ respectively.
The first order thermodynamical transition from the paramagnetic to the ferromagnetic phase happens on a curve lying between the two spinodal curves (not shown) and it is in general a non increasing function of $T$.
So the non-monotonicity of the curves $H_c(T)$ shown in Fig.~\ref{fig:phase_diag_severalC} should be taken as an indication that at low enough temperatures the phase transition becomes first order. According to the data presented in Fig.~\ref{fig:phase_diag_severalC} the first order transition takes place only for $C$ large enough.

\bibliographystyle{apsrev4-1}
\bibliography{myBiblio}

\end{document}